\documentstyle[12pt]{article}

\newcommand{\eqref}[1]{eq.(\ref{#1})}

\newcommand{\beq}{\begin{equation}}
\newcommand{\eeq}{\end{equation}}
\newcommand{\bl}[1]{\makebox[#1em]{}}

\newcommand{\pa}{\partial}

\newcommand{\ee}{\mbox{e}}

\def\rlx{\relax\leavevmode}

\def\IZ{\rlx\hbox{\sf Z\kern-.4em Z}}
\def\sIZ{\rlx\hbox{\scriptsize\sf Z\kern-.4em Z}}
\def\IN{\rlx\hbox{\rm I\kern-.18em N}}
\def\sIN{\rlx\hbox{\scriptsize\rm I\kern-.19em N}}
\def\IR{\rlx\hbox{\rm I\kern-.18em R}}
\def\IC{\rlx\hbox{\sf C\kern-.5em I }}

\newenvironment{eqabc}%
{\setcounter{enumi}{\value{equation}}%
\addtocounter{enumi}{1}%
\setcounter{equation}{0}%
\renewcommand{\theequation}{\theenumi\alph{equation}}%
\begin{eqnarray}}%
{\end{eqnarray}\setcounter{equation}{\value{enumi}}}%

\begin{document}
\hfill solv-int/9510006\\
\vspace{5pt}
\begin{center}
\begin{Large}
{\bf Toda Lattice Hierarchy and\\[3pt]
 Zamolodchikov's Conjecture}
\end{Large}\\[15pt]

Saburo Kakei\footnote{e-mail: kakei@mmm.t.u-tokyo.ac.jp}\\[5pt]
{\it Department of Mathematical Sciences, University of Tokyo,}\\
{\it Komaba 3-8-1, Meguro-ku, Tokyo 153, Japan}\\[15pt]

{\small\bf Abstract}\\[5pt]
\begin{minipage}[t]{14.5cm}
{\small\bl{2}In this letter, we show that certain Fredholm determinant
$D(\lambda;t)$, introduced by Zamolodchikov in his study of 2D
polymers, is a continuum limit of soliton solution for the Toda
lattice hierarchy with 2-periodic reduction condition.}\\[20pt]
\end{minipage}
\end{center}

In his recent study of 2D polymers on a cylinder, Zamolodchikov
conjectured that the quantity,
\beq
\phi(t) = \log\frac{D(\lambda;t)}{D(-\lambda;t)},\bl{2}
t=(t_{2n-1})_{n\in\sIZ},
\label{eq:phi}
\eeq
satisfies the differential equations of the integrated modified KdV
(mKdV) hierarchy\cite{Z}, and also satisfies the sinh-Gordon (shG)
equation.
In \eqref{eq:phi}, $D(\lambda;t)$ stands for the Fredholm determinant
$\det(I+\lambda K)$. Integral operator $K$ has a kernel of the form,
\beq
K(p,q) = \frac{e(p)e(q)}{p+q}
\exp\left(\sum_{n\in\sIZ}t_{2n-1} (p^{2n-1}+q^{2n-1})\right),
\label{eq:K}
\eeq
where the domain of integration for the operator is $(0,\infty)$.
The quantity $e(p)$ does not depend on $t_n$'s.
It should be remarked that our $t_n$ corresponds to $t_n/2$ of
refs.\cite{Z,TW}.
Independently, Bernard and LeClair showed that $\phi(t)$ solves the
shG equation\cite{BLC}.

Quite recently, Tracy and Widom proved that above $\phi(t)$ satisfies
both the mKdV hierarchy and the shG hierarchy\cite{TW}.
Their proof is based on the fact that the quantity $\phi(t)$ is
expressed in terms of Fredholm determinant.
In this letter, we will show that the Fredholm determinant
$D(\lambda;t)$ is nothing but a special case of
$\tau$-function for the Toda lattice hierarchy\cite{UT}.

As in the appendix D of ref.\cite{BLC},
we start with finite domain of integration $(0,\nu)$
for $\nu$ finite, rather than infinite domain $(0,\infty)$.
We then consider the quantity,
\beq
D_N(\lambda;t) =
\det\left(\delta_{jk} + \frac{\lambda\nu}{N} K(p_j, p_k)
\right)_{1\leq j,k\leq N},
\label{eq:DN}
\eeq
where $p_j = j\nu / N$ ($j=1,2,\ldots,N$).
As in the Fredholm theory, the Fredholm determinant $D(\lambda;t)
=\det(I+\lambda K)$ is recovered when we take the continuum limit
$N\rightarrow\infty$.
In the end, we take $\nu\rightarrow\infty$.
Expanding the r.h.s. of \eqref{eq:DN}, we come to the following
expression;
\beq
D_N(\lambda;t) =
\sum_{l=0}^{N}\left(\frac{\lambda\nu}{N}\right)^l
\sum_{j_1<\cdots<j_l}
\prod_{1\leq m<n\leq l}
\left( \frac{p_{j_m}-p_{j_n}}{p_{j_m}+p_{j_n}} \right)^2
\prod_{m=1}^{l}\frac{e(p_{j_m})^2}{2p_{j_m}}
\exp\left(\sum_{k\in\sIZ}2t_{2k-1} p_{j_m}^{2k-1}
\right) ,
\eeq
where we have used the Cauchy identity,
\[
\det\left(\frac{1}{x_i-y_j}\right)_{1\leq j,k\leq N} =
\frac{\prod_{1\leq i<j\leq n}(x_i-x_j)(y_j-y_i)}{
\prod_{1\leq i,j\leq n}(x_i-y_j)} .
\]
As we shall show below, this quantity $D_N(\lambda;t)$ is
the $\tau$-function which corresponds to the $N$-soliton solution of
the (2-periodic) Toda lattice hierarchy.

The Toda lattice hierarchy was formulated by Ueno and
Takasaki\cite{UT}(See also \cite{Kaji,Take}).
Here we briefly set up the notation.
Consider the following linear equations,
\begin{eqabc}
L(s,t;\ee^{\pa_s})w^{(\infty)}(s,t;\lambda) & = &
\lambda w^{(\infty)}(s,t;\lambda),\\
M(s,t;\ee^{\pa_s})w^{(0)}(s,t;\lambda) & = &
\lambda^{-1} w^{(0)}(s,t;\lambda),
\end{eqabc}%
where the difference operators $L(s;\ee^{\pa_s})$, $M(s;\ee^{\pa_s})$
are introduced by
\begin{eqabc}
L(s,t;\ee^{\pa_s}) & = & \sum_{-\infty<j\leq 1}b_j(s;t)\ee^{j\pa_s},
\bl{2.6}\mbox{with $b_1(s;t)=1$ for any $s$, $t$},\\
M(s,t;\ee^{\pa_s}) & = & \sum_{-1\leq j<+\infty}c_j(s;t)\ee^{j\pa_s},
\bl{2}\mbox{with $c_{-1}(s;t)\neq 0$ for any $s$, $t$},
\end{eqabc}%
and the wave functions $w^{(\infty)}$, $w^{(0)}$ by
\begin{eqabc}
w^{(\infty)}(s,t;\lambda) & = &
\left(\sum_{j=0}^{\infty}\hat{w}_j^{(\infty)}(s;t)\lambda^{-j}\right)
\lambda^s\exp\left( \sum_{j=1}^{\infty}t_j \lambda^j \right),\\
w^{(0)}(s,t;\lambda) & = &
\left(\sum_{j=0}^{\infty}\hat{w}_j^{(0)}(s;t)\lambda^j\right)
\lambda^s\exp\left( \sum_{j=1}^{\infty}t_{-j} \lambda^{-j} \right).
\end{eqabc}%
Time evolutions are defined through
\begin{eqabc}
\pa_n w
(s,t;\lambda) & = & B_n(s;\ee^{\pa_s})w(s,t;\lambda),\\
\pa_{-n} w
(s,t;\lambda) & = & C_n(s;\ee^{\pa_s})w(s,t;\lambda),
\end{eqabc}%
where $w$ stands for $w^{(\infty)}$ or $w^{(0)}$,
and $\pa_n = \frac{\pa}{\pa t_n}$.
The difference operators $B_n$ and $C_n$ are constructed from $L$ and
$M$ such that $B_n=(L^n)_+$, $C_n=(M^n)_-$ where $(\;\cdot\;)_+$
(or $(\;\cdot\;)_-$) denotes non-negative powers of $\ee^{\pa_s}$
(resp. negative).

Compatibility conditions of eq.(8) reduce to Zakharov-Shabat type
equations;
\begin{eqabc}
\pa_n B_m - \pa_m B_n & =& [ B_n, B_m ],\label{eq:ZSbb}\\
\pa_{-n} C_m - \pa_{-m} C_n & =& [ C_n, C_m],\label{eq:ZScc}\\
\pa_{-n} B_m - \pa_m C_n & =& [ C_n, B_m ],\label{eq:ZSbc}
\end{eqabc}%
for $m,n = 1,2,\ldots$.

Ueno and Takasaki proved the equivalence between Lax-type formulation
and the following {\it bilinear identity} (Th. 1.11 of \cite{UT});
\begin{eqnarray}
\lefteqn{
\sum_{j=0}^{\infty}p_{m+j}(-2y_+)p_j(\tilde{D}_+)
\exp\left(\sum_{j\neq 0}y_j D_j\right)
\tau(s+m+1;t)\cdot\tau(s;t)}\nonumber\\
 & = & \sum_{j=0}^{\infty}p_{-m+j}(-2y_-)p_j(\tilde{D}_-)
\exp\left(\sum_{j\neq 0}y_j D_j\right)
\tau(s+m;t)\cdot\tau(s+1;t),
\label{eq:BI}
\end{eqnarray}
for $s,m\in\IZ$, where we have set $t_{\pm}=(t_{\pm 1},t_{\pm 2},\ldots)$,
$y_{\pm}=(y_{\pm 1},y_{\pm 2},\ldots)$
and $\tilde{D}_{\pm} = (D_{\pm 1}, \frac{1}{2}D_{\pm
2},\frac{1}{3}D_{\pm 3},\ldots)$
stand for Hirota's bilinear operators;
\[
D_n f(t)\cdot g(t) =
\left.\frac{\pa}{\pa x}f(t_n +x)g(t_n -x)\right|_{x=0} .
\]
Polynomials $p_j(t)$ ($j=0,1,\ldots$) are introduced through
\[
\exp\left(\sum_{n=1}^{\infty}t_n \lambda^n \right) =
\sum_{n=1}^{\infty}p_j(t)\lambda^j.
\]
Expanding \eqref{eq:BI} w.r.t. $y_{\pm}$, we get Hirota
bilinear equations. For example, we have
\beq
\tau(s)\pa_1\pa_{-1}\tau(s) - \pa_1\tau(s)\pa_{-1}\tau(s)
+ \tau(s+1)\tau(s-1) = 0
\label{eq:BilToda}
\eeq
from coefficient of $y_{-1}$.

The wave functions  $w^{(\infty)}$, $w^{(0)}$ are constructed
in terms of $\tau$-function;
\begin{eqabc}
w^{(\infty)}(s,t;\lambda) & = &
\frac{\tau(s;t_+ -\epsilon(\lambda^{-1}),t_-)}{\tau(s;t)}
\lambda^s \exp\left( \sum_{j=1}^{\infty}t_j \lambda^j \right),\\
w^{(0)}(s,t;\lambda) & = &
\frac{\tau(s+1;t_+,t_- -\epsilon(\lambda))}{\tau(s;t)}
\lambda^s \exp\left( \sum_{j=1}^{\infty}t_{-j} \lambda^{-j} \right),
\end{eqabc}%
where $\epsilon(\lambda)=
(\lambda, \frac{1}{2}\lambda^2, \frac{1}{3}\lambda^3, \ldots)$.
{}From eqs.(5) and (12), we have
\begin{eqabc}
b_1(s;t) & = & \pa_1 \log\frac{\tau(s+1;t)}{\tau(s;t)},
\label{eq:BtoTAU}\\
c_0(s;t) & = & \frac{\tau(s+1;t)\tau(s-1;t)}{\tau(s;t)^2}.
\label{eq:CtoTAU}
\end{eqabc}%

Next we consider the reduction problem.
We impose 2-periodic condition on $\tau$-function, {\it i.e.}
$\tau(s+2;t) = \tau(s;t)$.
Under the 2-periodic condition, even time evolutions
$(t_{2n})_{n\neq 0}$ are suppressed.
Further, some conditions are imposed on $b_j$ and $c_j$\cite{Kaji}.
Here we list some of them;
\begin{eqnarray*}
b_1(s+1) & = & -b_1(s),\\
b_2(s+1) & = & -b_2(s)-b_1(s)^2,\\
c_0(s+1) & = & 1/c_0(s),\\
c_1(s+1) & = & - c_1(s),\\
b_1(s) & = & \frac{1}{2}\pa_1\log c_0(s),\\
\lefteqn{\cdots .}\nonumber
\end{eqnarray*}%
Under the conditions above, the Zakharov-Shabat type equations (9)
coincide with those of the mKdV and the shG hierarchy.
For example, we get
\setcounter{enumi}{\value{equation}}%
\addtocounter{enumi}{1}%
\setcounter{equation}{0}%
\renewcommand{\theequation}{\theenumi\alph{equation}}%
\beq
4\pa_3 b_1(0) = \pa_1^3 b_1(0) - 6 b_1(0)^2 \pa_1 b_1(0),
\label{eq:mKdV}
\eeq
from \eqref{eq:ZSbb} with $m=3$, $n=1$, and
\beq
\pa_1 \pa_{-1}\log c_0(0) = 2\left( c_0(0) -\frac{1}{c_0(0)}\right),
\label{eq:shG}
\eeq
\setcounter{equation}{\value{enumi}}%
\renewcommand{\theequation}{\arabic{equation}}%
from \eqref{eq:ZSbc} with $m=n=1$.

The $N$-soliton solution, which solves \eqref{eq:BI}, is constructed
as vacuum expectation value of fermion operators\cite{Take};
\setcounter{enumi}{\value{equation}}%
\addtocounter{enumi}{1}%
\setcounter{equation}{0}%
\renewcommand{\theequation}{\theenumi\alph{equation}}%
\beq
\tau(s;t) =
\langle s | \exp\left(\sum_{n>0}t_n H_n\right) g
\exp\left(-\sum_{n<0}t_{n} H_{n}\right) |s \rangle,
\eeq
where $g$ is an element of the Clifford group,
\beq
g = \exp\left(\sum_{j=1}^{N}a_j\psi(p_j)\psi^{*}(q_j)\right),
\label{eq:g}
\eeq
\setcounter{equation}{\value{enumi}}%
\renewcommand{\theequation}{\arabic{equation}}%
and $H_n$'s are boson operators obeying
$[H_m,H_n]=m\delta_{m+n,0}$.
For details of fermionic formulation, see ref.\cite{Take,JM}.

More explicitly, this $\tau$-function is of the form,
\setcounter{enumi}{\value{equation}}%
\addtocounter{enumi}{1}%
\setcounter{equation}{0}%
\renewcommand{\theequation}{\theenumi\alph{equation}}%
\beq
\tau_N(s;t)=\tau'_N(s;t)\exp\left(-\sum_{n>0}n t_n t_{-n}\right),
\label{eq:TtoTp}
\eeq
where $\tau'_N(s;t)$ is given by
\beq
\tau'_N(s;t) = \sum_{l=0}^{N}\sum_{j_1<\cdots<j_l}
  c_{j_1\cdots j_l} \prod_{m=1}^{l}a_{j_m}(s)
  \exp\left(\sum_{n\neq 0}
   t_n(p_{j_m}^n -q_{j_m}^n)\right),
\label{eq:Tp}
\eeq
\beq
a_j(s)=a_j \left(\frac{p_j}{q_j}\right)^{s}\frac{q_j}{p_j -
q_j},\bl{2}
c_{j_1\cdots j_l} = \prod_{1\leq m < n \leq l}
 \frac{(p_{j_m}-p_{j_n})(q_{j_m}-q_{j_n})}{(p_{j_m}-q_{j_n})
 (q_{j_m}-p_{j_n})}.
\eeq
\setcounter{equation}{\value{enumi}}%
\renewcommand{\theequation}{\arabic{equation}}%

Define an automorphism $\iota_2$ of fermion operators by
\[
\iota_2 (\psi(p)) = p^{-2}\psi(p), \bl{2}
\iota_2 (\psi^{*}(p)) = p^2 \psi^{*}(p).
\]
Imposing the condition $\iota_2 (g)=g$ on \eqref{eq:g}, it reduces to
$q_j = -p_j$ ($j=1,\ldots,N$). Further, this condition makes
$\tau$-function 2-periodic.
{}From Lie algebraic viewpoint, this reduction corresponds to the choice
of infinitesimal symmetry as the affine Lie algebra
$A_1^{(1)}$\cite{JM}.

If we set
\[
a_j = -\frac{\lambda\nu e(p_j)^2}{N p_j}, \bl{2}
p_j = -q_j =\frac{j\nu}{N},
\]
for $j=1,\ldots,N$, then $\tau'_N$ becomes 2-periodic, and
is identified to the quantity $D_N(\lambda;t)$;
\beq
\tau'_N(s=0;t) = D_N(\lambda;t), \bl{2}
\tau'_N(s=1;t) = D_N(-\lambda;t).
\label{eq:TtoD}
\eeq
Substituting eqs.(\ref{eq:TtoTp}), (\ref{eq:TtoD}) into eq.(13)
and taking the continuum limit $N,\nu\rightarrow\infty$,
we have the relation connecting $b_1$, $c_0$
to $\phi$ of \eqref{eq:phi};
\[
b_1(s=0;t) = -\pa_1\phi(t), \bl{2} \log c_0(s=0;t) = -2\phi(t).
\]
Substituting these into eqs.(14), we get
\begin{eqnarray*}
4\pa_3\phi & = & \pa_1^3\phi -2(\pa_1\phi)^3,\\
\pa_1\pa_{-1}\phi & = & 2\sinh 2\phi .
\end{eqnarray*}
The former is the integrated mKdV equation and the latter is the shG
equation. Other equations contained in the hierarchy is similarly
obtained from the Zakharov-Shabat equations (9).

In addition,
substituting eqs.(\ref{eq:TtoTp}), (\ref{eq:TtoD}) into
\eqref{eq:BilToda} and taking the limit $N,\nu\rightarrow\infty$,
we then have
\beq
D(\lambda;t)\pa_1\pa_{-1}D(\lambda;t)
- \pa_1 D(\lambda;t)\pa_{-1}D(\lambda;t)
= D(\lambda;t)^2 - D(-\lambda;t)^2
\eeq
This relation is also conjectured by Zamolodchikov (eq.(3.12) of
ref.\cite{Z}). From the viewpoint of the theory of $\tau$-function,
this relation is nothing but a member of the Pl\"uker relations.

We finally remark that ``continuum limit'' of soliton solution
for the KP hierarchy has been discussed by several authors.
For example, such type of $\tau$-function is referred to as
``general solution'' in ref.\cite{JM}. Their ``general solution'' is
of the form,
\[
\tau(s;t) = \langle s | \exp\left(\sum_{n>0}t_n H_n\right)
\exp\left(\int\!\int a(p,q)\psi(p)\psi^{*}(q)\mbox{d}p\mbox{d}q\right)
|s \rangle .
\]
In our case, if we take the limit $N ,\nu\rightarrow\infty$ on
eq.(15),
we get fermionic expression for the Fredholm determinant,
$D(\lambda;t)$,
\begin{eqnarray}
\lefteqn{D(\lambda;t) = \exp\left(\sum_{n>0}n t_n t_{-n}\right)}
\nonumber\\
 & & \times\langle s | \exp\left(\sum_{n>0}t_n H_n\right)
\exp\left(\int a(p)\psi(p)\psi^{*}(-p)\mbox{d}p\right)
\exp\left(-\sum_{n<0}t_{n} H_{n}\right) |s \rangle .
\label{eq:tauD}
\end{eqnarray}
Though the r.h.s of \eqref{eq:tauD} is rather formal,
the limit $N\rightarrow\infty$ makes sense since the l.h.s is
well-defined so far as $K$ of \eqref{eq:K} belongs to trace-class.

We also note that Mulase has discussed ``continuum soliton solution''
on his study on matrix integrals\cite{Mu}.
We hope that ``continuum limit'' of soliton solution would give
another viewpoint to understand the structure of integrable field
theory.

\section*{Acknowledgement}
\bl{1}
The author acknowledges Professor Junkichi Satsuma
for his critical reading of the manuscript.

\end{document}